\documentclass{article}
\usepackage{spconf,amsmath,graphicx,hyperref}
\usepackage{booktabs}
\usepackage{multirow}
\usepackage{arydshln}
\usepackage[table]{xcolor}
\usepackage{colortbl}
\usepackage{acronym}
\usepackage{graphicx}
\usepackage{enumitem}
\usepackage{hyperref}
\usepackage{booktabs}
\usepackage{makecell}
\usepackage{siunitx}
\usepackage{pifont}
\usepackage{makecell}
\usepackage{pgfplots}
\usepackage{amssymb}
\pgfplotsset{compat=1.18} % Use the latest compatibility mode
\usetikzlibrary{patterns} % For hatch patterns, if needed
\usepackage{tikz}
\newcommand{\cmark}{\ding{51}}
\newcommand{\xmark}{\ding{55}}

\definecolor{darkorange}{RGB}{200,100,0}
\definecolor{lightorange}{RGB}{235,150,60}

\newacro{dl}[DL]{deep learning}
\newacro{dnn}[DNN]{deep neural network}
\newacroplural{dnn}[DNNs]{deep neural networks}
\newacro{cnn}[CNN]{convolutional neural network}
\newacroplural{cnn}[CNNs]{convolutional neural networks}
\newacro{rnn}[RNN]{recurrent neural network}
\newacroplural{rnn}[RNNs]{recurrent neural networks}
\newacro{gan}[GAN]{generative adversarial network}
\newacroplural{gan}[GANs]{generative adversarial networks}
\newacro{vae}[VAE]{variational autoencoder}
\newacroplural{vae}[VAEs]{variational autoencoders}
\newacro{cvae}[CVAE]{conditional VAE}
\newacro{mlp}[MLP]{multilayer perceptron}
\newacroplural{mlp}[MLPs]{multilayer perceptrons}
\newacro{nb}[NB]{narrowband}
\newacro{wb}[WB]{wideband}
\newacro{bwe}[BWE]{bandwidth extension}
\newacro{ssr}[SSR]{speech super-resolution}
\newacro{asr}[SR]{audio super-resolution}
\newacro{se}[SE]{speech enhancement}
\newacro{lr}[LR]{low-resolution}
\newacro{hr}[HR]{high-resolution}
\newacro{lf}[LF]{low-frequency}
\newacro{hf}[HF]{high-frequency}
\newacro{stft}[STFT]{short-time Fourier transform}
\newacro{lps}[LPS]{log-power spectrum}
\newacro{mfcc}[MFCC]{mel-frequency cepstral coefficient}
\newacro{lsf}[LSF]{line spectral frequency}
\newacro{bpvc}[BPVC]{band-pass voicing coefficient}
\newacro{lpc}[LPC]{linear predictive coding}
\newacro{dct}[DCT]{discrete cosine transform}
\newacro{acf}[AFC]{autocorrelation function coefficient}
\newacro{zcr}[ZCR]{zero-crossing rate}
\newacro{gbrbm}[GBRBM]{Gaussian-Bernoulli restricted Boltzmann machine}
\newacro{pqmf}[PQMF]{Pseudo Quadrature Mirror Filter}
\newacro{mdct}[MDCT]{modified discrete cosine transform}
\newacro{hmm}[HMM]{hidden Markov model}
\newacroplural{hmm}[HMMs]{hidden Markov models}
\newacroplural{gmm}[GMMs]{Gaussian mixture models}
\newacro{ar}[AR]{autoregressive}
\newacro{llm}[LLM]{large language model}
\newacroplural{llm}[LLMs]{large language models}
\newacro{lalm}[LALM]{large audio-language model}
\newacroplural{lam}[LAMs]{large audio-language models}

\newacro{rvq}[RVQ]{residual vector quantization}
\newacro{ode}[ODE]{ordinary differential equation}
\newacroplural{CNF}[CNFs]{continuous normalizing flows}
\newacroplural{sde}[SDEs]{stochastic differential equations}
\newacro{snr}[SNR]{Signal-to-Noise Ratio}
\newacro{lsd}[LSD]{Log Spectral Distance}

\newacro{mmse}[MMSE]{minimum mean-square error}
\newacro{istft}[iSTFT]{inverse STFT}
\newacro{mse}[MSE]{mean squared error}
\newacro{mae}[MAE]{mean absolute error}
\newacro{sisdr}[SI-SDR]{scale-invariant signal-to-distortion ratio}
\newacro{nll}[NLL]{negative log-likelihood}
\newacro{ce}[CE]{cross-entropy}
\newacro{mrstft}[MR-STFT]{multi-resolution STFT}
\newacro{fm}[FM]{feature matching}
\newacro{lb}[LB]{lower-band}
\newacro{ub}[UB]{upper-band}
\newacro{bl}[BL]{band-limited}
\newacro{bb}[BB]{broadband}
\newacro{rbms}[RBMs]{restricted Boltzmann machines}
\newacro{gmm}[GMM]{Gaussian mixture model}
\newacro{lstm}[LSTM]{long short-term memory}
\newacro{blstm}[BLSTM]{bidirectional LSTM}
\newacro{lpcc}[LPCC]{linear predictive cepstral coefficient}
\newacro{bidilconv}[Bi-DilConv]{bidirectional dilated convolution}
\newacro{stfc}[STFC]{short-time Fourier convolution}
\newacro{bsft}[BSFT]{bandwidth spectral feature transform}
\newacro{udm}[UDM]{unconditional diffusion model}
\newacro{sgm}[SGM]{score-based generative model}
\newacroplural{sgm}[SGMs]{score-based generative models}
\newacro{cfm}[CFM]{conditional FM}
\newacro{fm}[FM]{flow matching}
\newacro{ldm}[LDM]{latent diffusion model}
\newacro{kl}[KL]{Kullback-Leibler}
\newacro{hrnn}[HRNN]{hierarchical RNN}
\newacro{lsgan}[LSGAN]{least-squares GAN}
\newacro{nac}[NAC]{neural audio
codec}
\newacro{1d}[1-D]{one-dimensional}
\newacro{2d}[2-D]{two-dimensional} 
\title{P2Flow: Phoneme-aware Progressive Flow Matching for Extreme Speech Super-Resolution}
\name{
\shortstack{
Ningyuan Yang$^{1*}$,
Yize Li$^{2*}$,
Pu Zhao$^{2}$,
Diego A. Cuji$^{1}$,
Kanad Sarkar$^{3}$,\\
Ryan M. Corey$^{4}$,
Xue Lin$^{2}$,
Andrew C. Singer$^{1}$
}
\thanks{$^*$Equal contribution.
Code is available 
\href{https://github.com/ningyuan33/P2Flow}{here}.
}
}

\address{
$^{1}$Stony Brook University, NY, US \;
$^{2}$Northeastern University, MA, US \;
$^{3}$UIUC, IL, US \;
$^{4}$UIC, IL, US
}

\begin{document}
\ninept
\maketitle
\begin{abstract}
Generative models have recently demonstrated considerable promise in speech super-resolution (SSR). Nevertheless, the majority of existing work has concentrated on standard or versatile SSR configurations, leaving the extreme setting with severely limited spectral inputs largely unexplored. In this regime, current approaches exhibit marked performance degradation, underscoring the need for dedicated solutions.
To bridge this gap, we introduce P2Flow, a phoneme-aware progressive flow matching (FM) framework designed for extreme SSR with three main strategies. First, our model leverages phonetic information to reconstruct missing spectral components. Furthermore, it employs a progressive architectural design that hierarchically restores distinct frequency regions. Finally, we incorporate post-training of the vocoder to enhance overall waveform fidelity.
Extensive experiments are conducted on the TIMIT and VCTK datasets under both 1 kHz to 16 kHz and 2 kHz to 16 kHz settings, demonstrating that P2Flow yields state-of-the-art results across multiple evaluation metrics.
\end{abstract}
\section{Introduction}
\label{sec:intro}

\Ac{ssr} \cite{yang2026survey, eskimez2019speech}, historically termed as speech bandwidth extension~\cite{li2015deep}, aims to reconstruct \ac{hr} speech from a \ac{lr} input. Early  approaches~\cite{kuleshov2017audio,lim2018time,birnbaum2019temporal,rakotonirina2021self,wang2021towards,nguyen2022tunet} have employed discriminative models to learn direct \ac{lr}-to-\ac{hr} mappings, yet often yielded over-smoothed spectra with limited \ac{hf} details. More recent generative paradigms
~\cite{ho2020denoising,lipman2022flow,zhou2024denoising} have demonstrated superior \ac{hf} restoration: \acp{gan}~\cite{mandel2023aero,hauret2023eben,lu2024towards,lee2025wave,zhang2025vm} improve perceptual realism via adversarial training; diffusion models~\cite{lee2021nu,han2022nu,yu2023conditioning,liu2024audiosr,fang2025vector} achieve high-fidelity reconstruction through iterative denoising; \ac{fm}~\cite{yun2025flowhigh,choi2026universr,im2026saga,hsieh2026towards,zhang2026codecflow} offers efficient sampling via continuous-time dynamics; and Schrödinger bridges~\cite{kong2025a2sb,li2025bridge,li2026audio} learn stochastic transport between \ac{lr} and \ac{hr} distributions.

\begin{figure}[t]
\centering
\includegraphics[width=\columnwidth]{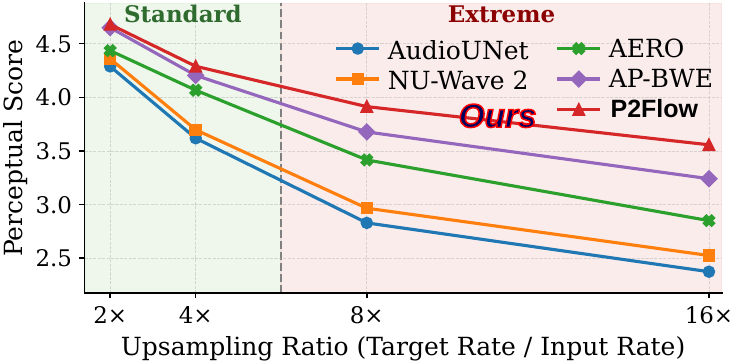}
\vspace{-5mm}
\caption{Perceptual Score for \ac{ssr} on VCTK with target rate 16~kHz.}
\label{fig:visqol_input_rate}
\vspace{-5mm}
\end{figure}

However, existing research has predominantly focused on standard \Ac{ssr} settings, with moderately reduced input sampling rates (e.g., input$\rightarrow$output: 8~kHz$\rightarrow$16~kHz), or versatile settings (e.g., arbitrary input rates$\rightarrow$48~kHz)~\cite{han2022nu,liu2024audiosr,choi2026universr}. In contrast, the problem of extreme \Ac{ssr} (e.g., 1 or 2~kHz$\rightarrow$16~kHz), wherein only severely limited spectral information is available, has received considerably less attention, despite its practical significance in applications such as bone-conduction sensing~\cite{hauret2023eben,li2023two}.
To systematically investigate the limitations of current approaches, we first evaluate several representative \Ac{ssr} methods on VCTK~\cite{yamagishi2019vctk} with a fixed 16~kHz target rate under varying input rates. As illustrated in Fig.~\ref{fig:visqol_input_rate}, reconstruction performance degrades notably as upsampling ratio increases, underscoring the substantial difficulty of \ac{hf} restoration when acoustic and phonetic cues are largely insufficient for extreme \Ac{ssr}.

\begin{figure*}[t]
\centering
\includegraphics[width=0.95\textwidth]{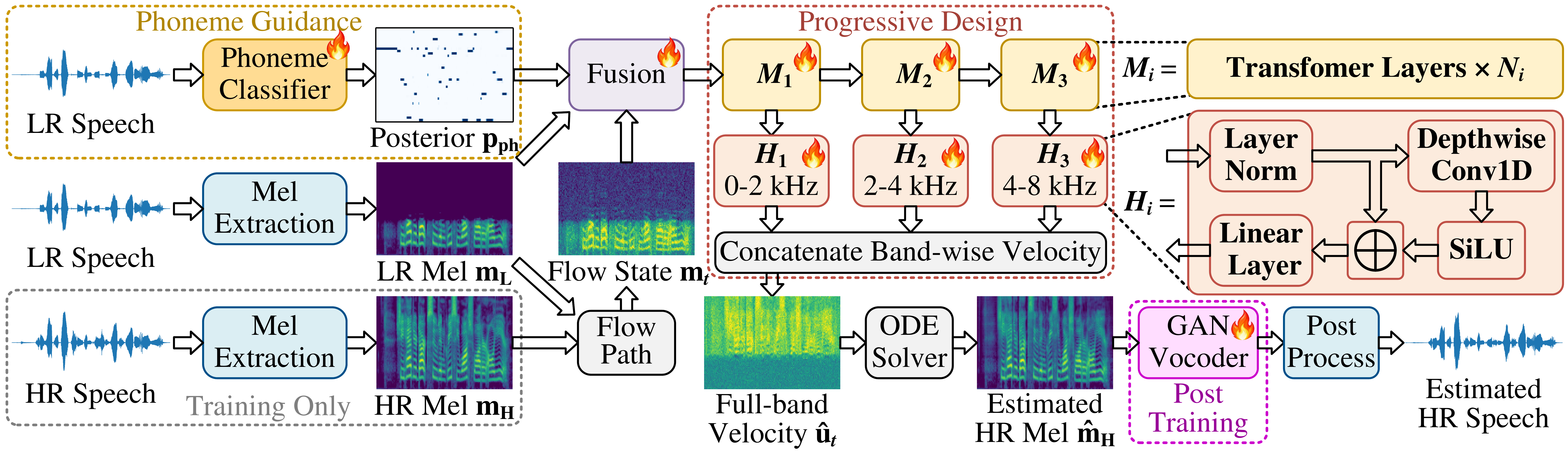}
\vspace{-4mm}
\caption{Overview of the proposed P2Flow framework for \ac{hr} speech restoration from \ac{lr} speech. Training follows three steps: (1) phoneme classifier training, (2) \ac{fm} training with the other modules frozen, and (3) vocoder post-training using estimated mel-spectrograms.}
\label{fig:p2flow}
\vspace{-3mm}
\end{figure*}

To address these limitations, we propose \textbf{\textit{P2Flow}}, a \textbf{P}honeme-aware \textbf{P}rogressive \textbf{Flow} matching framework for extreme \Ac{ssr}.
Building upon few-step FM~\cite{yun2025flowhigh}, we introduce three key strategies: phoneme guidance, progressive design, and vocoder post-training. (i) Specifically, recognizing that phoneme identity encodes strong cues about harmonic structure and spectral envelope~\cite{hou2020speaker,poli2024improving}, we first integrate a HuBERT-based phoneme classifier into the model. It conditions the FM process on frame-level phoneme posterior distributions to guide \ac{hf} reconstruction. (ii) In the extreme SSR situation, direct recovery of the largely missing spectral content is particularly difficult. Motivated by image super-resolution~\cite{wang2018fully,zamir2021multi}, we therefore adopt a progressive modeling scheme that decomposes the reconstruction task across frequency bands, enabling HF estimation to leverage lower-frequency representations hierarchically. (iii) To mitigate errors from mel estimation to waveform synthesis, we additionally post-train the vocoder with adversarial learning. Thorough experiments on the TIMIT~\cite{garofolo1988getting} and VCTK~\cite{yamagishi2019vctk} datasets show that P2Flow consistently outperforms representative baselines in spectral fidelity, perceptual speech quality, and intelligibility. Ablation studies further confirm the individual contribution of each proposed component.
Overall, P2Flow has demonstrated superiority over other approaches in both standard and extreme settings, as presented in Fig.~\ref{fig:visqol_input_rate}. 
Our contributions are summarized as follows:
\begin{itemize}
\vspace{-1mm}
    \item We identify the overlooked performance degradation of existing \ac{ssr} methods under extreme band-limited settings.
    \vspace{-1mm}
    \item We introduce P2Flow, a phoneme-aware progressive \ac{fm} framework for extreme \ac{ssr}, comprising phoneme guidance, progressive frequency modeling, and vocoder post-training.
    \vspace{-1mm}
    \item Comprehensive experiments on TIMIT and VCTK demonstrate state-of-the-art performance and validate each strategy.
\end{itemize}

\section{Flow Matching Preliminaries}
\label{sec:prelim}

Let $\mathbf{x}\in\mathbb{R}^{d}$ denote a generic data representation. 
\Ac{fm}~\cite{lipman2022flow,yun2025flowhigh} defines a probability path $p_t$, $t\in[0,1]$, that transports a source distribution $p_0$ to a target distribution $p_1$. A flow state $\mathbf{x}_t\sim p_t$ evolves according to a time-dependent velocity field $v_\theta$ as
$\frac{d\mathbf{x}_t}{dt}=v_\theta(\mathbf{x}_t,t)$.
\Ac{fm} learns this velocity field such that the induced trajectories follow the prescribed probability path. For a Gaussian path with mean $\boldsymbol{\mu}_t$ and standard deviation $\sigma_t$, the flow state is sampled as
$
\mathbf{x}_t=\boldsymbol{\mu}_t+\sigma_t\boldsymbol{\epsilon},
$
where $\boldsymbol{\epsilon}\sim\mathcal{N}(\mathbf{0},\mathbf{I})$.
Differentiating with respect to $t$ gives the target velocity
$
\mathbf{u}_t=\dot{\boldsymbol{\mu}}_t+\dot{\sigma}_t\boldsymbol{\epsilon}.
$
The predicted velocity field $v_\theta(\mathbf{x}_t,t)$ is trained to match this target by
\begin{equation}
\mathcal{L}_{\mathrm{FM}}
=
\mathbb{E}_{t,\boldsymbol{\epsilon}}
\left[
\left\|
\mathbf{u}_t-v_\theta(\mathbf{x}_t,t)
\right\|_2^2
\right].
\end{equation}
At inference, a sample $\mathbf{x}_0\sim p_0$ is transported toward the target distribution by integrating the learned ordinary differential equation (ODE) $\frac{d\mathbf{x}_t}{dt}=v_\theta(\mathbf{x}_t,t)$ from $t=0$ to $t=1$, yielding $\mathbf{x}_1\sim p_1$.

\section{Proposed Method}
\label{sec:method}

P2Flow consists of three complementary components for extreme \ac{ssr}: (i) a phoneme-aware classifier for lingual conditioning; (ii) progressive hierarchical design in \ac{fm} to simplify wide-band reconstruction, and (iii) vocoder post-training to reduce mel-to-waveform reconstruction errors. The system pipeline is illustrated in Fig.~\ref{fig:p2flow}.

\begin{table*}[t]
\centering
\caption{16~kHz extreme SSR performance on TIMIT. Lower ($\downarrow$) is better for LSD metrics; higher ($\uparrow$) is desirable for ViSQOL and STOI.}
\label{tab:main_results1}

\begingroup
\setlength{\tabcolsep}{3.5pt}
\renewcommand{\arraystretch}{0.6}

\resizebox{\textwidth}{!}{%
\begin{tabular}{l|ccccc|ccccc}
\toprule
\multirow{2}{*}{Model} 
& \multicolumn{5}{c|}{\raisebox{0.3ex}{1~kHz $\rightarrow$ 16~kHz}} 
& \multicolumn{5}{c}{\raisebox{0.3ex}{2~kHz $\rightarrow$ 16~kHz}} \\
\cline{2-11}
& \raisebox{-1.15ex}{LSD $\downarrow$} & \raisebox{-1.15ex}{LSD-LF $\downarrow$} & \raisebox{-1.15ex}{LSD-HF $\downarrow$} & \raisebox{-1.15ex}{ViSQOL $\uparrow$} & \raisebox{-1.15ex}{STOI $\uparrow$} & \raisebox{-1.15ex}{LSD $\downarrow$} & \raisebox{-1.15ex}{LSD-LF $\downarrow$} & \raisebox{-1.15ex}{LSD-HF $\downarrow$} & \raisebox{-1.15ex}{ViSQOL $\uparrow$} & \raisebox{-1.15ex}{STOI $\uparrow$} \\ \midrule
Input & 4.550 & 0.723 & 4.694 & 1.498 & 0.663 & 4.234 & 0.195 & 4.525 & 2.355 & 0.791 \\
\cmidrule(lr){1-11}
AudioUNet~\cite{kuleshov2017audio} & 2.549 & 0.561 & 2.626 & 2.440 & 0.752 & 2.436 & 0.242 & 2.600 & 3.036 & 0.834 \\
\cmidrule(lr){1-11}
NU-Wave 2~\cite{han2022nu}  & 2.129 & 0.482 & 2.193 & 2.458 & 0.695 & 1.690 & 0.354 & 1.799 & 3.049 & 0.797 \\
UDM+~\cite{yu2023conditioning} & 1.656 & 0.355 & 1.707 & 2.174 & 0.629 & 1.456 & 0.254 & 1.552 & 2.910 & 0.782 \\
\cmidrule(lr){1-11}
AERO~\cite{mandel2023aero} & 1.440 & 0.732 & 1.476 & 2.778 & 0.755 & 1.285 & 0.557 & 1.359 & 3.268 & 0.847 \\
EBEN~\cite{hauret2023eben} & 1.126 & 0.569 & 1.152 & 3.139 & 0.796 & 1.057 & 0.584 & 1.105 & 3.603 & 0.871 \\
AP-BWE~\cite{lu2024towards} & 1.069 & \textbf{0.343} & 1.110 & 3.155 & 0.800 & 0.998 & 0.274 & 1.060 & 3.647 & 0.875 \\
\cmidrule(lr){1-11}
FLowHigh~\cite{yun2025flowhigh} & 1.105 & 0.724 & 1.124 & 3.149 & 0.799 & 1.017 & 0.195 & 1.084 & 3.712 & 0.873 \\
UniverSR~\cite{choi2026universr} & 1.216 & 0.365 & 1.251 & 2.841 & 0.761 & 1.114 & 0.263 & 1.186 & 3.325 & 0.853 \\
\rowcolor{gray!30}
\textbf{P2Flow (ours)} & \textbf{1.026} & 0.723 & \textbf{1.041} & \textbf{3.434} & \textbf{0.849} & \textbf{0.935} & \textbf{0.195} & \textbf{0.997} & \textbf{3.954} & \textbf{0.908} \\
\bottomrule
\end{tabular}%
}
\endgroup
\vspace{-1mm}
\end{table*}

\begin{figure*}[htbp]
\centering
\includegraphics[width=\textwidth]{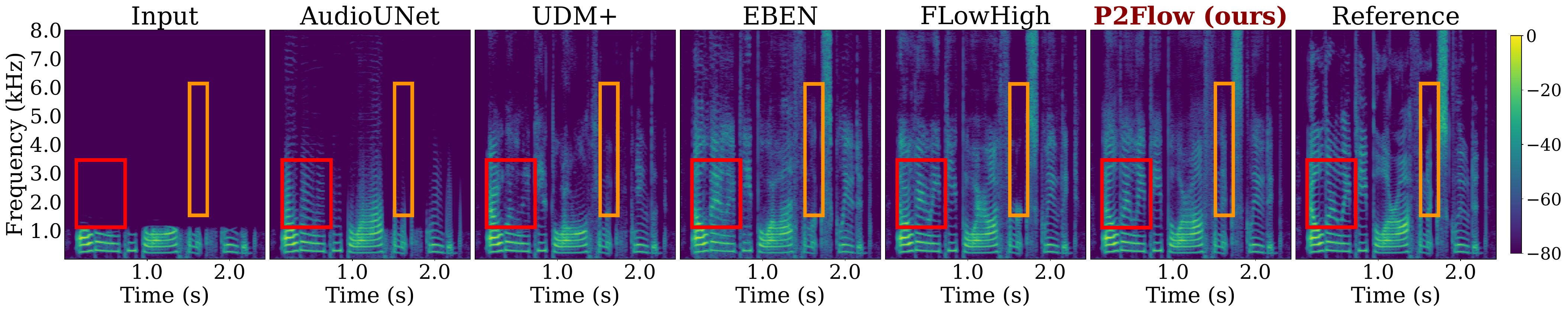}
\vspace{-6mm}
\caption{Spectrogram comparison on TIMIT under the 2~kHz$\rightarrow$16~kHz \ac{ssr} setting. Colored boxes highlight noticeable spectral differences.}
\label{fig:timeline}
\vspace{-4mm}
\end{figure*}

\subsection{Phoneme Classifier}
To provide linguistic guidance for \ac{hf} restoration~\cite{poli2024improving}, we first train a phoneme classifier consisting of a HuBERT encoder~\cite{hsu2021hubert} and a linear head for frame-level phoneme prediction. Following the standard TIMIT 61-to-39 phone mapping~\cite{lee1989speaker}, we merge acoustically similar variants into 39 classes.
The classifier is trained on speech degraded identically to the \ac{ssr} input. Since the HuBERT encoder expects 16~kHz waveforms, the \ac{lr} signal $\mathbf{s}_{\mathrm{L}}$ is first resampled to 16~kHz and normalized, yielding $\tilde{\mathbf{s}}_{\mathrm{L}}$ before feature extraction.
We temporally interpolate the HuBERT features from a 20~ms stride to match the 16~ms mel-spectrogram frame stride, then apply the linear head to obtain frame-level phoneme posterior probabilities by
\vspace{-1mm}
\begin{equation}
\mathbf{p}_{\mathrm{ph}}
=
\mathrm{softmax}\!\left(
C_{\mathrm{ph}}\!\left(
\mathcal{I}\!\left(Enc(\tilde{\mathbf{s}}_{\mathrm{L}})\right)
\right)
\right)
\in \mathbb{R}^{B\times 39 \times T},
\end{equation}
where $Enc(\cdot)$ denotes the HuBERT encoder, $\mathcal{I}(\cdot)$ represents temporal interpolation, $C_{\mathrm{ph}}(\cdot)$ is the linear classification head, $B$ is the batch size, and $T$ is the number of mel-spectrogram frames. The classifier is then frozen during \ac{ssr} training, while $\mathbf{p}_{\mathrm{ph}}$ provides frame-level linguistic conditioning to the progressive \ac{fm} model.

\subsection{Progressive Modeling}

Extreme \ac{ssr} requires  reconstructing a largely missing frequency range, 
which is particularly challenging.
We therefore decompose this difficult mapping into progressive frequency-region modeling, allowing later modules to build on increasingly refined representations.
Let $\mathbf{m}_{\mathrm{L}},\mathbf{m}_{\mathrm{H}}\in\mathbb{R}^{B\times F\times T}$ denote the LR and HR mel-spectrograms, where $F$ is the number of mel-frequency bins.
We define  $\boldsymbol{\mu}_t=t\mathbf{m}_{\mathrm{H}}+(1-t)\mathbf{m}_{\mathrm{L}}$ and $\sigma_t=1-(1-\sigma_{\min})t$, and sample the flow state as $\mathbf{m}_t=\boldsymbol{\mu}_t+\sigma_t\boldsymbol{\epsilon}$, where $\boldsymbol{\epsilon}\sim\mathcal{N}(\mathbf{0},\mathbf{I})$. The corresponding target velocity is $\mathbf{u}_t=(\mathbf{m}_{\mathrm{H}}-\mathbf{m}_{\mathrm{L}})-(1-\sigma_{\min})\boldsymbol{\epsilon}$. The flow state, LR condition, and phoneme posterior are then concatenated and linearly projected as $\mathbf{h}_0=W_{\mathrm{in}}[\mathbf{m}_t;\mathbf{m}_{\mathrm{L}};\mathbf{p}_{\mathrm{ph}}]$. As shown in Fig.~\ref{fig:p2flow}, our design is based on a few-step \Ac{fm}~\cite{yun2025flowhigh} with progressive phoneme awareness in the mel-spectrogram domain.

Three sequential modules hierarchically refine the embedded representation as $\mathbf{h}_k=M_k(\mathbf{h}_{k-1})$, $k\in\{1,2,3\}$. We use the 4/2/2-layer Transformer for $M_1$, $M_2$, and $M_3$, respectively, with a hidden dimension of 1024. Each layer in the Transformer block consists of time-conditioned adaptive RMS normalization, followed by a 16-head self-attention (64-dimensional heads) incorporating query-key RMS normalization and rotary positional embeddings, and then a GEGLU feed-forward network with residual connections. The reversed pyramid design allocates more capacity to $M_1$ since the first module initially infers spectral structure from the most limited information, while subsequent modules benefit from progressively abundant features. Each $M_k$ is followed by a lightweight velocity head $H_k$ that predicts the target velocity in the lower-frequency (0--2~kHz), middle-frequency (2--4~kHz), or upper-frequency (4--8~kHz) mel region as $\hat{\mathbf{u}}_t^{(k)}=H_k(\mathbf{h}_k)$. Specifically, each head first applies layer normalization, then a depthwise temporal convolution followed by a SiLU activation with a residual connection, and finally a linear projection, as shown in Fig.~\ref{fig:p2flow}. The three predictions form the full 80-bin velocity field, optimized via 
\vspace{-2mm}
\begin{equation}
\mathcal{L}_{\mathrm{FM}}=\sum_{k=1}^{3}\left\|\hat{\mathbf{u}}_t^{(k)}-\mathbf{u}_t^{(k)}\right\|_2^2. 
\vspace{-0.5mm}
\end{equation}
During inference, we initialize $\mathbf{m}_0=\mathbf{m}_{\mathrm{L}}+\sigma\boldsymbol{\epsilon}$ and integrate $\mathbf{v}_{\theta}(\mathbf{m}_t,t,\mathbf{m}_{\mathrm{L}},\mathbf{p}_{\mathrm{ph}})$ from $t=0$ to $1$ to estimate the \ac{hr} mel-spectrogram $\hat{\mathbf{m}}_{\mathrm{H}}$.

\subsection{Vocoder Post-training}
For 16~kHz extreme \ac{ssr}, we adopt the SpeechBrain HiFi-GAN vocoder~\cite{kong2020hifi} for mel-to-waveform synthesis.
Since it is pretrained on natural mel-spectrograms, a distribution mismatch arises when decoding mel-spectrogram estimates from the \ac{fm} model. We therefore post-train the generator on fixed mel estimates while freezing the \ac{fm} model. A discriminator is jointly trained, while the generator is optimized using standard GAN-vocoder objectives~\cite{kong2020hifi,lee2022bigvgan,yamamoto2020parallel,yang2021multi}, including adversarial, feature-matching, and multi-resolution short-time Fourier transform (MR-STFT) losses, as
% \vspace{-1mm}
\begin{equation}
\mathcal{L}_{G}=\lambda_{\mathrm{adv}}\mathcal{L}_{\mathrm{adv}}+\lambda_{\mathrm{feat}}\mathcal{L}_{\mathrm{feat}}+\lambda_{\mathrm{STFT}}\mathcal{L}_{\mathrm{MR\text{-}STFT}}.
\end{equation}
Specifically, the adversarial and feature-matching losses are given as
\begin{equation}
\mathcal{L}_{\mathrm{adv}}
=
\sum_k
\left\|
D_k(\hat{\mathbf{y}})-1
\right\|_2^2.
\end{equation}

\begin{equation}
\mathcal{L}_{\mathrm{feat}}
=
\frac{1}{N}
\sum_{k,l}
\left\|
D_k^{(l)}(\hat{\mathbf{y}})
-
D_k^{(l)}(\mathbf{y})
\right\|_1.
\end{equation}
% \vspace{-1mm}
where $\hat{\mathbf{y}}$ and $\mathbf{y}$ are the generated and target waveforms, $D_k$ is the $k$-th sub-discriminator, $D_k^{(l)}$ its $l$-th intermediate feature map, and $N$ the total number of feature maps. The MR-STFT loss is
\begin{equation}
\mathcal{L}_{\mathrm{MR\text{-}STFT}}
=
\frac{1}{R}
\sum_{r=1}^{R}
\left(
\mathcal{L}_{\mathrm{SC}}^{(r)}
+
\mathcal{L}_{\mathrm{MAG}}^{(r)}
\right),
\end{equation}
where $\mathcal{L}_{\mathrm{SC}}^{(r)}$ and $\mathcal{L}_{\mathrm{MAG}}^{(r)}$ denote the spectral-convergence and log-magnitude losses at resolution $r$, respectively. They are defined as
\begin{equation}
\mathcal{L}_{\mathrm{SC}}^{(r)}
=
\frac{
\left\|
S_r(\mathbf{y})-S_r(\hat{\mathbf{y}})
\right\|_F
}{
\left\|
S_r(\mathbf{y})
\right\|_F
},
\end{equation}
\begin{equation}
\mathcal{L}_{\mathrm{MAG}}^{(r)}
=
\left\|
\log S_r(\mathbf{y})
-
\log S_r(\hat{\mathbf{y}})
\right\|_1,
\end{equation}
where $R$ is the number of STFT resolutions, $S_r(\cdot)$ denotes the STFT magnitude at resolution $r$, and $\|\cdot\|_F$ is the Frobenius norm. 
The adversarial and feature-matching losses improve waveform realism and training stability, while the MR-STFT loss preserves spectral structure across multiple resolutions, together reducing errors from mel estimation to vocoder waveform synthesis.

\begin{table*}[t]
\centering
\caption{16~kHz extreme SSR performance on VCTK. Lower ($\downarrow$) is better for LSD metrics; higher ($\uparrow$) is desirable for ViSQOL and STOI.}
\label{tab:main_results}

\begingroup
\setlength{\tabcolsep}{3.5pt}
\renewcommand{\arraystretch}{0.6}

\resizebox{\textwidth}{!}{%
\begin{tabular}{l|ccccc|ccccc}
\toprule
\multirow{2}{*}{Model} 
& \multicolumn{5}{c|}{\raisebox{0.3ex}{1~kHz $\rightarrow$ 16~kHz}} 
& \multicolumn{5}{c}{\raisebox{0.3ex}{2~kHz $\rightarrow$ 16~kHz}} \\
\cline{2-11}
& \raisebox{-1.15ex}{LSD $\downarrow$} & \raisebox{-1.15ex}{LSD-LF $\downarrow$} & \raisebox{-1.15ex}{LSD-HF $\downarrow$} & \raisebox{-1.15ex}{ViSQOL $\uparrow$} & \raisebox{-1.15ex}{STOI $\uparrow$} & \raisebox{-1.15ex}{LSD $\downarrow$} & \raisebox{-1.15ex}{LSD-LF $\downarrow$} & \raisebox{-1.15ex}{LSD-HF $\downarrow$} & \raisebox{-1.15ex}{ViSQOL $\uparrow$} & \raisebox{-1.15ex}{STOI $\uparrow$} \\ 
\midrule
Input & 4.002 & 0.273 & 4.132 & 1.779 & 0.685 & 3.778 & 0.185 & 4.037 & 2.296 & 0.808 \\
\cmidrule(lr){1-11}
AudioUNet~\cite{kuleshov2017audio} & 2.447 & \textbf{0.228} & 2.525 & 2.374 & 0.774 & 2.323 & 0.197 & 2.482 & 2.829 & 0.831 \\
% \cmidrule(lr){1-11}
NU-Wave 2~\cite{han2022nu}  & 1.917 & 0.286 & 1.977 & 2.525 & 0.725 & 1.681 & 0.283 & 1.793 & 2.966 & 0.814 \\
AP-BWE~\cite{lu2024towards} & 1.172 & 0.251 & 1.208 & 3.241 & 0.805 & 1.065 & 0.231 & 1.135 & 3.677 & 0.876 \\
% \cmidrule(lr){1-11}
FLowHigh~\cite{yun2025flowhigh} & 1.144 & 0.273 & 1.179 & 3.391 & 0.826 & 1.054 & 0.185 & 1.124 & 3.752 & 0.884 \\
\rowcolor{gray!30}
\textbf{P2Flow (ours)} & \textbf{1.056} & 0.274 & \textbf{1.088} & \textbf{3.557} & \textbf{0.853} & \textbf{0.979} & \textbf{0.185} & \textbf{1.043} & \textbf{3.913} & \textbf{0.906} \\
\bottomrule
\end{tabular}%
}
\endgroup
\vspace{-4mm}
\end{table*}

\section{Experiments}
\label{sec:exp}

\subsection{Experimental Setup}

\subsubsection{Training}
% \xhdr{Training}
All models are trained and evaluated under identical 1~kHz$\rightarrow$16~kHz and 2~kHz$\rightarrow$16~kHz \ac{ssr} settings on TIMIT~\cite{garofolo1988getting} using the official train/test split and VCTK~\cite{yamagishi2019vctk} using a 100/8-speaker train/test split. VCTK phoneme labels are generated with Montreal Forced Aligner (MFA)~\cite{mcauliffe2017montreal} and mapped to the TIMIT 39-phone set. For each mel frame, we take its temporal midpoint and assign the phoneme label whose aligned time interval contains that point. LR inputs are generated by low-pass filtering the 16~kHz ground truth with a Chebyshev Type-I filter and then downsampling to 1 or 2~kHz. We utilize 80-bin mel-spectrograms with a 1024-sample frame length and 256-sample hop length. The phoneme classifier is trained for 20 epochs, while the \ac{fm} model is trained for 400k iterations with a batch size of 128 and a learning rate of $3\times10^{-4}$. The vocoder is fine-tuned for 20 epochs with $(\lambda_{\mathrm{adv}},\lambda_{\mathrm{feat}},\lambda_{\mathrm{STFT}})=(1,10,1)$ using three MR-STFT resolutions specified by (FFT size, hop size, window size): $(512,128,512)$, $(1024,256,1024)$, and $(2048,512,2048)$. 
The \ac{fm} model is evaluated using 5 sampling steps as the default. We follow the same post-processing strategy in FLowHigh~\cite{yun2025flowhigh} to preserve the reliable \ac{lf} spectrum. All experiments are conducted on a single NVIDIA RTX A6000 GPU. 

\subsubsection{Evaluation}
Overall spectral distortion is quantified using log-spectral distance (LSD)~\cite{liu2024audiosr} (lower is better). We further report LSD-LF and LSD-HF over the preserved \ac{lf} and missing \ac{hf} regions. We adopt the virtual speech quality objective listener (ViSQOL)~\cite{chinen2020visqol} and short-time objective intelligibility (STOI)~\cite{taal2010short} to evaluate perceptual speech quality and intelligibility, respectively, where higher scores are better.

\subsubsection{Baseline}
% \xhdr{Baselines}
We compare P2Flow with a broad set of representative \ac{ssr} methods, including AudioUNet~\cite{kuleshov2017audio} as a discriminative model; AERO~\cite{mandel2023aero}, EBEN~\cite{hauret2023eben}, and AP-BWE~\cite{lu2024towards} as \ac{gan}-based approaches; NU-Wave 2~\cite{han2022nu} and UDM+~\cite{yu2023conditioning} as diffusion-based baselines; and FLowHigh~\cite{yun2025flowhigh} and UniverSR~\cite{choi2026universr} as \ac{fm}-based methods. For all baselines, we follow the official implementations.

\subsection{Main Results}
Table~\ref{tab:main_results1} compares P2Flow with diverse baselines on TIMIT under two extreme settings. P2Flow achieves the best overall performance across both settings, with the lowest LSD/LSD-HF and the highest ViSQOL and STOI. The pronounced gains demonstrate P2Flow's effectiveness in recovering severely missing \ac{hf} spectral content while improving perceptual speech  quality and intelligibility. Furthermore, Fig.~\ref{fig:timeline} provides a spectrogram comparison on TIMIT. Compared with FLowHigh~\cite{yun2025flowhigh}, P2Flow more closely matches the ground-truth spectral structure: the red boxes show more accurate spectral-gap reconstruction, while the orange boxes
highlight \ac{hf} components recovered by P2Flow but largely missed by FLowHigh. Similar improvements are observed on VCTK in Table~\ref{tab:main_results}, demonstrating consistent performance gains across datasets. Additionally, P2Flow even maintains its superior performance under the 4~kHz$\rightarrow$16~kHz and 8~kHz$\rightarrow$16~kHz settings, as shown in Fig.~\ref{fig:visqol_input_rate}. More importantly, its advantage persists across various sampling conditions, confirming its general effectiveness for both standard and extreme SSR.

\begin{table}[htbp]
\vspace{-4mm}
\centering
\caption{Ablation study of P2Flow on TIMIT (2~kHz$\rightarrow$16~kHz).}
\label{tab:ablation1}

\setlength{\tabcolsep}{2.0pt}
\renewcommand{\arraystretch}{0.50}

\resizebox{\columnwidth}{!}{
\begin{tabular}{ccc|ccc}
\toprule
Progressive & Phoneme & Post-train
& LSD $\downarrow$
& ViSQOL $\uparrow$
& STOI $\uparrow$ \\
\midrule
\xmark & \xmark & \xmark & 0.989 & 3.831 & 0.888 \\
\cmark & \xmark & \xmark & 0.975 & 3.848 & 0.892 \\
\cmark & \xmark & \cmark & 0.962 & 3.856 & 0.893 \\
\cmark & \cmark & \xmark & 0.950 & 3.943 & 0.907 \\
\cmark & \cmark & \cmark & \textbf{0.935} & \textbf{3.954} & \textbf{0.908} \\
\bottomrule
\end{tabular}%
}
\vspace{-6mm}
\end{table}

\subsection{Ablation Study}
To further verify the effectiveness of each proposed component, we conduct an ablation study on TIMIT under the 2~kHz$\rightarrow$16~kHz setting. In Table~\ref{tab:ablation1}, progressive modeling consistently improves all metrics, while phoneme conditioning yields the most distinct gains in perceptual quality and intelligibility. Additionally, vocoder post-training reduces spectral distortion and improves overall quality. The optimal performance is achieved by combining all proposed designs. 

Moreover, we investigate the layer allocation across the three Transformer modules $M_k$, $k\in\{1,2,3\}$. As shown in Table~\ref{tab:pyramid_design}, the 4/2/2 design achieves the best performance, suggesting that devoting more layers to the earlier module effectively mitigates the greater ambiguity inherent in lower-frequency spectral reconstruction.

In addition, we report ViSQOL via different sampling steps, achieving scores of 3.946 (1‑step), 3.954 (5‑step), and 3.954 (10‑step). It demonstrates that P2Flow remains highly consistent across inference step counts, thereby underscoring the favorable trade‑off between quality and computational cost. 

\begin{table}[htbp]
\vspace{-5mm}
\centering
\caption{Comparison of Transformer layer allocation across the $M_1$, $M_2$, and $M_3$ modules in P2Flow on TIMIT (2~kHz$\rightarrow$16~kHz).}
\label{tab:pyramid_design}

\setlength{\tabcolsep}{4.5pt}
\renewcommand{\arraystretch}{0.50}

\resizebox{\columnwidth}{!}{%
\begin{tabular}{ccc|ccc}
\toprule
$M_1$ & $M_2$ & $M_3$
& LSD $\downarrow$
& ViSQOL $\uparrow$
& STOI $\uparrow$ \\
\midrule
2-layer & 2-layer & 4-layer & 0.948 & 3.925 & 0.906 \\
2-layer & 4-layer & 2-layer & 0.941 & 3.947 & 0.907 \\
\textbf{4-layer} & \textbf{2-layer} & \textbf{2-layer}
& \textbf{0.935} & \textbf{3.954} & \textbf{0.908} \\
\bottomrule
\end{tabular}%
}
\vspace{-6mm}
\end{table}

\subsection{Phoneme Analysis}
\begin{figure}[h]
\vspace{-2.5mm}
\centering
\includegraphics[width=1\columnwidth]{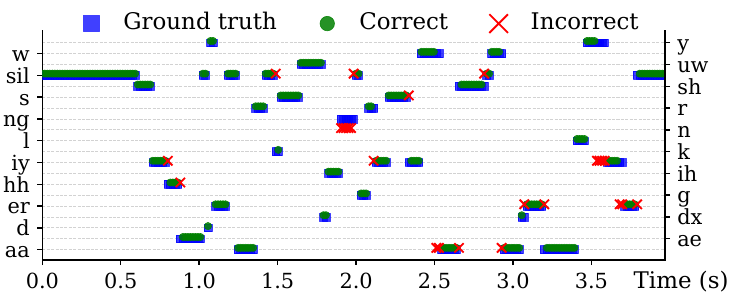}
\vspace{-5mm}
\caption{An example of temporal phoneme predictions on TIMIT.}
\label{fig:phoneme_prediction}
\vspace{-4.0mm}
\end{figure}

We examine the reliability of phoneme guidance under severely band-limited inputs. The phoneme classifier achieves 82.55\% and 75.64\% test accuracy for 2~kHz and 1~kHz inputs on TIMIT, respectively. It indicates that substantial phonetic information remains recoverable even under extreme bandwidth reduction. As shown in Fig.~\ref{fig:phoneme_prediction}, most errors occur between acoustically similar phones, such as $\{\textit{ng},\textit{n}\}$ and $\{\textit{iy},\textit{y}\}$, or near phoneme boundaries where transitional cues make classification ambiguous. Rather than relying on hard phoneme labels, our P2Flow instead uses posterior probabilities to retain uncertainty among plausible phones, providing robust phonetic guidance for reconstructing missing \ac{hf} content.

\section{Conclusion}

This work investigates the underexplored problem of extreme \ac{ssr} and proposes P2Flow, a phoneme-aware progressive \ac{fm} framework. P2Flow combines phoneme guidance, progressive spectral generation, and vocoder post-training to improve reconstruction under severe bandwidth constraints. Evaluated on TIMIT and VCTK, P2Flow surpasses competitive baselines by a clear margin in spectral, perceptual, and intelligibility metrics. Moreover, our ablation and phoneme-based investigations confirm the individual contribution and necessity of each proposed component. Future work will investigate the extension of P2Flow to more diverse and realistic bandwidth-degradation scenarios.

\vfill\pagebreak
% \newpage
\clearpage

\bibliographystyle{IEEEbib}
% \small
\bibliography{refs}

@inproceedings{li2015deep,
  title={A deep neural network approach to speech bandwidth expansion},
  author={Li, Kehuang and others},
  booktitle={IEEE ICASSP},
  year={2015}
}

@inproceedings{mandel2023aero,
  title={Aero: Audio super resolution in the spectral domain},
  author={Mandel, Moshe and others},
  booktitle={IEEE ICASSP},
  year={2023},
}

@article{kuleshov2017audio,
  title={Audio super resolution using neural networks},
  author={Kuleshov, Volodymyr and Enam, S Zayd and Ermon, Stefano},
  journal={arXiv preprint arXiv:1708.00853},
  year={2017}
}

@inproceedings{han2022nu,
  title={NU-Wave 2: A general neural audio upsampling model for various sampling rates},
  author={Han, Seungu and Lee, Junhyeok},
  booktitle={Proc. Interspeech},
  year={2022}
}

@inproceedings{yu2023conditioning,
  title={Conditioning and sampling in variational diffusion models for speech super-resolution},
  author={Yu, Chin-Yun and Yeh, Sung-Lin and Fazekas, György and others},
  booktitle={IEEE ICASSP},
  year={2023},
}

@inproceedings{hauret2023eben,
  title={EBEN: Extreme bandwidth extension network applied to speech signals captured with noise-resilient body-conduction microphones},
  author={Hauret, Julien and others},
  booktitle={IEEE ICASSP},
  year={2023},
}

@article{lu2024towards,
  title={Towards high-quality and efficient speech bandwidth extension with parallel amplitude and phase prediction},
  author={Ye-Xin Lu and others},
  journal={IEEE/ACM TASLP},
  year={2024},
}

@inproceedings{yun2025flowhigh,
  title={Flowhigh: Towards efficient and high-quality audio super-resolution with single-step flow matching},
  author={Yun, Jun-Hak and others},
  booktitle={IEEE ICASSP},
  year={2025},
}

@inproceedings{choi2026universr,
  title={Universr: Unified and versatile audio super-resolution via vocoder-free flow matching},
  author={Woongjib Choi and Lee, Sangmin and others},
  booktitle={IEEE ICASSP},
  year={2026},
}

@article{yang2026survey,
  title={A survey of advancing audio super-resolution and bandwidth extension from discriminative to generative models},
  author={Yang, Ningyuan and Li, Yize and others},
  journal={arXiv preprint arXiv:2605.16681},
  year={2026}
}

@inproceedings{birnbaum2019temporal,
  title={Temporal FiLM: Capturing long-range sequence dependencies with feature-wise modulations.},
  author={Birnbaum, Sawyer and Kuleshov, Volodymyr and others},
  booktitle={NeurIPS},
  year={2019}
}

@inproceedings{lim2018time,
  title={Time-frequency networks for audio super-resolution},
  author={Lim, Teck Yian and Yeh, Raymond A. and others},
  booktitle={IEEE ICASSP},
  year={2018},
}

@inproceedings{rakotonirina2021self,
  title={Self-attention for audio super-resolution},
  author={Rakotonirina, Nathana{\"e}l Carraz},
  booktitle={IEEE MLSP},
  year={2021},
}

@article{wang2021towards,
  title={Towards robust speech super-resolution},
  author={Wang, Heming and Wang, DeLiang},
  journal={IEEE/ACM TASLP},
  year={2021},
}

@inproceedings{nguyen2022tunet,
  title={Tunet: A block-online bandwidth extension model based on transformers and self-supervised pretraining},
  author={Nguyen, Viet-Anh and others},
  booktitle={IEEE ICASSP},
  year={2022},
}

@inproceedings{lee2025wave,
  title={Wave-u-mamba: an end-to-end framework for high-quality and efficient speech super resolution},
  author={Lee, Yongjoon and Kim, Chanwoo},
  booktitle={IEEE ICASSP},
  year={2025},
}

@article{zhang2025vm,
  title={VM-ASR: A lightweight dual-stream U-net model for efficient audio super-resolution},
  author={Zhang, Ting-Wei and Ruan, Shanq-Jang},
  journal={IEEE/ACM TASLP},
  year={2025},
}

@inproceedings{eskimez2019speech,
  title={Speech super resolution generative adversarial network},
  author={Eskimez, Sefik Emre and others},
  booktitle={IEEE ICASSP},
  year={2019},
}

@inproceedings{lee2021nu,
  title={Nu-wave: A diffusion probabilistic model for neural audio upsampling},
  author={Lee, Junhyeok and Han, Seungu},
  booktitle={Proc. Interspeech},
  year={2021}
}

@inproceedings{liu2024audiosr,
  title={AudioSR: Versatile audio super-resolution at scale},
  author={Liu, Haohe and Chen, Ke and others},
  booktitle={IEEE ICASSP},
  year={2024}
}

@inproceedings{fang2025vector,
  title={Vector quantized diffusion model based speech bandwidth extension},
  author={Fang, Yuan and others},
  booktitle={IEEE ICASSP},
  year={2025},
}

@inproceedings{im2026saga,
  title={Saga-sr: Semantically and acoustically guided audio super-resolution},
  author={Im, Jaekwon and others},
  booktitle={IEEE ICASSP},
  year={2026},
}

@inproceedings{hsieh2026towards,
  title={Towards real-time generative speech restoration with flow-matching},
  author={Hsieh, Tsun-An and others},
  booktitle={IEEE ICASSP},
  year={2026},
}

@article{zhang2026codecflow,
  title={Codecflow: Efficient bandwidth extension via conditional flow matching in neural codec latent space},
  author={Zhang, Bowen and others},
  journal={arXiv preprint arXiv:2603.02022},
  year={2026}
}

@inproceedings{li2025bridge,
  title={Bridge-sr: Schr{\"o}dinger bridge for efficient sr},
  author={Li, Chang and Chen, Zehua and others},
  booktitle={IEEE ICASSP},
  year={2025},
}

@inproceedings{li2026audio,
  title={Audio super-resolution with latent bridge models},
  author={Li, Chang and Chen, Zehua and others},
  booktitle={NeurIPS},
  year={2026}
}

@article{kong2025a2sb,
  title={A2sb: Audio-to-audio schrodinger bridges},
  author={Kong, Zhifeng and others},
  journal={arXiv preprint arXiv:2501.11311},
  year={2025}
}

@inproceedings{poli2024improving,
  title={Improving spoken language modeling with phoneme classification: A simple fine-tuning approach},
  author={Poli, Maxime and Chemla, Emmanuel and others},
  booktitle={EMNLP},
  year={2024}
}

@article{hsu2021hubert,
  title={Hubert: Self-supervised speech representation learning by masked prediction of hidden units},
  author={Hsu, Wei-Ning and Bolte, Benjamin and others},
  journal={IEEE/ACM TASLP},
  year={2021},
}

@article{lee1989speaker,
  title={Speaker-independent phone recognition using hidden Markov models},
  author={Lee, K-F and others},
  journal={IEEE/ACM TASLP},
  year={1989},
}

@inproceedings{kong2020hifi,
  title={Hifi-gan: Generative adversarial networks for efficient and high fidelity speech synthesis},
  author={Kong, Jungil and Kim, Jaehyeon and Bae, Jaekyoung},
  booktitle={NeurIPS},
  year={2020}
}

@inproceedings{yamamoto2020parallel,
  title={Parallel WaveGAN: A fast waveform generation model based on generative adversarial networks with multi-resolution spectrogram},
  author={Yamamoto, Ryuichi and Song, Eunwoo and Kim, Jae-Min},
  booktitle={IEEE ICASSP},
  year={2020},
}

@inproceedings{lipman2022flow,
  title={Flow matching for generative modeling},
  author={Yaron Lipman and Ricky T. Q. Chen and others},
  booktitle={ICLR},
  year={2023}
}

@inproceedings{mcauliffe2017montreal,
  title={Montreal forced aligner: Trainable text-speech alignment using kaldi},
  author={McAuliffe, Michael and others},
  booktitle={Proc. Interspeech},
  year={2017}
}

@article{li2023two,
  title={A two-stage approach to quality restoration of bone-conducted speech},
  author={Li, Changtao and others},
  journal={IEEE/ACM TASLP},
  year={2023},
}

@inproceedings{wang2018fully,
  title={A fully progressive approach to single-image super-resolution},
  author={Wang, Yifan and others},
  booktitle={CVPRW},
  year={2018},
}

@inproceedings{zamir2021multi,
  title={Multi-stage progressive image restoration},
  author={Zamir, Syed Waqas and others},
  booktitle={CVPR},
  year={2021},
}

@inproceedings{hou2020speaker,
  title={Speaker and phoneme-aware speech bandwidth extension with residual dual-path network},
  author={Hou, Nana and Xu, Chenglin and Pham, Van Tung and others},
  booktitle={Proc. Interspeech},
  year={2020}
}

@inproceedings{ho2020denoising,
  title={Denoising diffusion probabilistic models},
  author={Ho, Jonathan and Jain, Ajay and Abbeel, Pieter},
  booktitle={NeurIPS},
  year={2020}
}

@inproceedings{zhou2024denoising,
  title={Denoising diffusion bridge models},
  author={Zhou, Linqi and Lou, Aaron and Khanna, Samar and Ermon, Stefano},
  booktitle={ICLR},
  year={2024}
}

@inproceedings{lee2022bigvgan,
  title={Bigvgan: A universal neural vocoder with large-scale training},
  author={Lee, Sang-gil and Ping, Wei and others},
  booktitle={ICLR},
  year={2023}
}

@article{garofolo1988getting,
  title={Getting started with the DARPA TIMIT CD-ROM: An acoustic phonetic continuous speech database},
  author={Garofolo, John S and Lamel, Lori F and others},
  journal={NIST, Gaithersburgh, MD},
  year={1988}
}

@misc{yamagishi2019vctk,
  author       = {Yamagishi, Junichi and Veaux, Christophe and MacDonald, Kirsten},
  title        = {CSTR VCTK Corpus: English Multi-speaker Corpus for CSTR Voice Cloning Toolkit (version 0.92)},
  year         = {2019},
  publisher    = {University of Edinburgh, The Centre for Speech Technology Research (CSTR)},
}

@inproceedings{chinen2020visqol,
  title={ViSQOL v3: An open source production ready objective speech and audio metric},
  author={Chinen, Michael and Lim, Felicia SC and others},
  booktitle={IEEE QoMEX},
  year={2020},
}

@inproceedings{taal2010short,
  title={A short-time objective intelligibility measure for time-frequency weighted noisy speech},
  author={Taal, Cees H and Hendriks, Richard C and others},
  booktitle={IEEE ICASSP},
  year={2010},
}

@inproceedings{yang2021multi,
  title={Multi-band melgan: Faster waveform generation for high-quality text-to-speech},
  author={Yang, Geng and others},
  booktitle={IEEE SLT},
  year={2021},
}

\end{document}